\documentclass[aps,prd,amsmath,two column,amssymb,showpacs]{revtex4}
\usepackage{amssymb}
\usepackage{txfonts}
\usepackage{mathbbol}
\usepackage{amsfonts}
\usepackage{mathrsfs}
\usepackage{epsfig,bm,dcolumn}
\usepackage{graphicx}
\usepackage{color}
\usepackage{amsmath}
\usepackage{dcolumn}
\usepackage{overpic}
\usepackage{slashed}
\usepackage{hyperref}

\begin{document}
\title{Competition between magnetic catalysis effect and chiral rotation effect}

\author{Lingxiao Wang$^1$ and Gaoqing Cao$^{2,3}$}
\affiliation{1 Department of Physics, Tsinghua University and Collaborative Innovation Center of Quantum Matter, Beijing 100084, China\\
2 School of Physics and Astronomy, Sun Yat-Sen University, Guangzhou 510275, China\\
3 Department of Physics and Center for Particle Physics and Field Theory, Fudan University, Shanghai 200433, China.}
\date{\today}

\begin{abstract}
In this work, we explore the competition between magnetic catalysis effect and chiral rotation effect in a general parallel electromagnetic field within the effective Nambu--Jona-Lasinio model. For a given electric field $E$ at zero temperature, the mass gap shows three different features with respect to an increasing magnetic field $B$: increasing monotonically, decreasing after increasing and decreasing monotonically. By making use of strong magnetic field approximation, we illuminate that this is due to the competition between catalysis effect and chiral rotation effect induced both by the magnetic field, and a critical electric field $\sqrt{eE_c}=86.4~{\rm MeV}$ is found beyond which the mass gap will eventually decrease at large $B$. As only large magnetic field is relevant for the derivation, the critical electric field does not depend on the temperature $T$ or chemical potential $\mu$.
\end{abstract}

\maketitle
\section{Introduction}
Recently, experimentalists receive two most important goals in the field of high energy nuclear physics, that is to fix the critical end point (CEP) in $T-\mu$ phase diagram~\cite{Luo:2017faz} and discover chiral magnetic effect (CME)~\cite{Liao:2014ava,Kharzeev:2015znc,Huang:2015oca} in quantum chromodynamics (QCD) systems through relativistic heavy ion collisions (HICs). The corresponding tough technical challenges are mainly creating a large baryon density system at high temperature for CEP and eliminating the smearing background flow signals for CME, respectively. The critical end point is mainly related to chiral symmetry breaking and restoration~\cite{Klevansky:1989vi} with respect to temperature $T$ and baryon chemical potential $\mu$, to which a lot of efforts have been devoted in the theoretical aspects: In the first principal lattice QCD (LQCD) simulations, the chiral transition was found to be a crossover at small $\mu$ and the critical temperature is around $T_c=155~{\rm MeV}$~\cite{Borsanyi:2010cj,Borsanyi:2013bia,Bazavov:2014pvz}. For larger $\mu$, chiral effective models usually predicted a first-order transition when neglecting inhomogeneous phases, such as Nambu--Jona-Lasinio (NJL) model~\cite{Zhuang:1994dw}, Polyakov--Nambu--Jona-Lasinio (PNJL) model\cite{Fukushima:2008wg}, linear sigma model\cite{Bilic:1997sh}, MIT bag model \cite{Gupta:2011wh}, quark-meson model~\cite{Herbst:2013ail} holographic QCD model~\cite{Li:2017ple}, etc.. More recently, since the discovery of intriguing inverse magnetic catalysis effect (IMCE) around the critical temperature in LQCD simulations~\cite{Bali:2011qj,Bali:2012zg,Bruckmann:2013oba}, the research on chiral symmetry breaking and restoration enters a new era and a lot of works have been done to explore the QCD properties in the background of external magnetic field~\cite{Fukushima:2012kc,Kojo:2012js,Chao:2013qpa,Yu:2014sla,Cao:2014uva,Ferrer:2014qka,Cao:2015xja,Cao:2016fby}. CME is also a magnetic field related phenomenon, which means an electric current induced along the magnetic field when the chiral charge density is nonzero~\cite{Kharzeev:2007jp,Fukushima:2008xe}. Besides, other chiral anomaly related topics were also widely studied, such as chiral separation effect, chiral magnetic wave, chiral electric separation effect, chiral vortical effect and magnetovorticity effect~\cite{Son:2004tq,Metlitski:2005pr,Erdmenger:2008rm,Son:2009tf,Gorbar:2009bm,Kharzeev:2010gd,Burnier:2011bf,Banerjee:2008th,Jiang:2015cva,Hattori:2016njk}.

From the previous introduction, if in any case the spontaneous chiral symmetry breaking and restoration can be explored together with chiral anomaly effects, probably in the presence of electromagnetic (EM) field~\cite{Miransky:2015ava}, the discussion might be very interesting. In our previous work, the effect of triangle anomaly to chiral symmetry breaking and restoration was explored in the presence of parallel EM field~\cite{Cao:2015cka}. By varying solely the second Lorentz invariant $\sqrt{I_2}=E=B$, we found that while pion condensate can be developed due to the chiral rotation effect (CRE) caused by EM chiral anomaly, the chiral symmetry always tends to be restored with $I_2$ as had been found in Ref.~\cite{Babansky:1997zh}. However, according to the studies at zero temperature~\cite{Bali:2011qj,Bali:2012zg,Gusynin:1994re,Gusynin:1994xp,Gusynin:1995nb,Miransky:2015ava}, usually the
 magnetic catalysis effect (MCE) was assumed due to the dimensional reduction and high degeneracy induced by the magnetic field. Thus, if the magnetic field $B$ is varied at a fixed electric field $E$, the competition between MCE and CRE might take place for chiral condensate. More concretely, it's just the competition between the first Lorentz invariant $I_1\equiv{{\bf B}^2-{\bf E}^2}$ with MCE and the second Lorentz invariant $I_2\equiv{\bf E\cdot B}$ with CRE. This is the main motivation of this work.

The paper is organized as the following: In Sec.\ref{NJL}, we give the formalism of chiral symmetry breaking and restoration in the presence of a general parallel EM field in Nambu--Jona-Lasinio (NJL) model, where the thermodynamic potential and gap equations at zero temperature are derived in Sec.\ref{T0} and nontrivial extensions to the finite temperature and chemical potential case are given in Sec.\ref{Tmu}. In Sec.\ref{calculations}, numerical results are presented with some detailed discussions. Finally, we give the conclusions and prospectives in Sec.\ref{conclusions}.
%%%%%%%%%%%%%%%%%%%%%%%%%%%%%%%%%%%%%%%%%%%%%%%%%%%%%%%%%%%%%%%%%%%%%%%%%%%%%%%%%%%%%%%%%%%%%%%%%%%%%%%%%%%%%%%%%%%%%%%%%%%%%%%%%%%%%%%%%%%%%%%%%%%%%%%%%%%
\section{Formalism in Nambu--Jona-Lasinio model}\label{NJL}
\subsection{Thermodynamic potential and gap equations}\label{T0}
For the case with a general parallel electromagnetic (EM) field, that is, $\bf{E\parallel B}$, the initial Lagrangian of Nambu--Jona-Lasinio model~\cite{Nambu:1961fr,Klevansky:1989vi} can be generalized to the following form in Euclidean space~\cite{Cao:2015cka}:
\begin{eqnarray}
{\cal L}_{\rm NJL}=\bar\psi(i\slashed{D}-m_0)\psi+G[(\bar\psi\psi)^2+(\bar\psi i\gamma_5{\tau}\psi)^2],
\end{eqnarray}
where $\psi=(u,d)^T$ represents the two-flavor quark field, $m_0$ is the current quark mass, $G$ is the four-fermion coupling constant, and
\begin{eqnarray}
D_\mu=\partial_\mu-iQA_\mu,
\end{eqnarray}
is the covariant derivative with quark charge matrix $Q={\rm diag}(2/3,-1/3)e$ in flavor space and $A_\mu$ representing a parallel EM field (without lose of generality, we can set $A_\mu=(iEz,0,-Bx,0)$). As has been illuminated in Ref.~\cite{Cao:2015cka}, the case with $I_2\neq0$ usually favors $\pi^0$ superfluid due to the triangle anomaly; thus, the neutral $\pi^0$ condensate should also be taken into account for the general study.

In mean field approximation, an effective action can be derived by using Hubbard-Stratonovich transformation $\sigma=-2G\bar\psi\psi$ and ${\boldsymbol \pi}=-2G\bar\psi i\gamma_5{\boldsymbol \tau}\psi$:
\begin{eqnarray}\label{action}
{\cal S}_{NJL} &=&\int{d^4x}{(m-m_0+\hat\sigma)^2+(\pi^0+\hat\pi^0)^2+\hat\pi_1^2+\hat\pi_2^2\over 4G}\nonumber\\
&&-{\rm Tr}\ln\bigg[i{\slashed D}-m-i\gamma_5{\tau_3}{\pi^0}-\hat\sigma-i\gamma_5{\boldsymbol \tau}\hat{\boldsymbol \pi}\bigg],
\end{eqnarray}
where the expectation values have already been taken out in the neutral sector, that is, $\sigma=m-m_0+\hat\sigma$ and $\pi_3=\pi^0+\hat\pi^0$. Then, the thermodynamic potential can be obtained by setting all the fluctuation fields $\hat{\sigma}$ and $\hat{\boldsymbol \pi}$ to zero:
\begin{eqnarray}
\Omega(m,\pi^0)={(m-m_0)^2+(\pi^0)^2\over 4G}-{{\rm Tr}\ln\left[i{\slashed D}-m-i\gamma_5{\tau_3}{\pi^0}\right]\over V_{3+1}}
\end{eqnarray}
with $V_{3+1}$ the space-time volume. And the coupled gap equations follow directly from the extremal conditions $\partial\Omega/\partial m=0$ and $\partial\Omega/\partial \pi^0=0$ as
\begin{eqnarray}
{m-m_0\over 2G}-{1\over V_{3+1}}\text {Tr}\;{\cal S}_{A}(x)&=&0,\label{gap1}\\
{\pi^0\over 2G}-{1\over V_{3+1}}\text {Tr}\;{\cal S}_{A}(x)i\gamma^5\tau_3&=&0,\label{gap2}
\end{eqnarray}
where the fermion propagator in the constant EM field is ${\cal S}_{A}(x)=-\left[i{\slashed D}-m-i\gamma^5\tau_3\pi^0\right]^{-1}$
which actually decouples for $u$ and $d$ quarks. 

Here, we take $u$ quark for example to show how the explicit form of fermion propagator can be derived in this case by following Schwinger's method~\cite{Schwinger:1951nm}. In Euclidean space, the propagator of $u$ quark can be formally expressed as:
\begin{eqnarray}
{\cal S}_{u}&=&\left[\gamma\Pi_{u}+m+i\gamma^5\pi^0\right]^{-1}\nonumber\\
&=&(-\gamma\Pi_{u}+m-i\gamma^5\pi^0)i\int_0^\infty ds~e^{-i[M^2-(\gamma\Pi_u)^2]s},
\end{eqnarray}
where $\Pi_u^\mu=-iD_u^\mu$ is the conjugate energy-momentum operator and we have defined the chiral mass $M\equiv(m^2+(\pi^0)^2)^{1/2}$. Then the current algebra gives the normal interaction between quark and external EM field as
\begin{eqnarray}
iq_{u} {\rm Tr}\; \gamma\delta A {\cal S}_{u}&=&-{\rm Tr}\;\delta(\gamma\Pi_{u})\gamma\Pi_{u}\int_0^\infty ds~e^{-i[M^2-(\gamma\Pi_u)^2]s}\nonumber\\
&=&\delta\Big[{i\over2}\int_0^\infty dss^{-1}{\rm Tr}e^{-i[M^2-(\gamma\Pi_{u})^2]s}\Big],
\end{eqnarray}
and we can easily identify an effective Lagrangian function in coordinate space:
\begin{eqnarray}
{\cal L}^{(1)}(x)&=&{i\over2}\int_0^\infty dss^{-1}e^{-iM^2s}{\rm tr}\langle x|U(s)|x\rangle,\\
U(s)&=&e^{-i{\cal H}s},
\end{eqnarray}
where the effective Hamiltonian in the proper-time evolution operator $U(s)$ is:
\begin{eqnarray}
{\cal H}=-(\gamma\Pi_{u})^2=\Pi_{u}^2-{1\over2}q_{u}\sigma_{\mu\nu}F_{\mu\nu},
\end{eqnarray}
with $\sigma_{\mu\nu}={i\over2}[\gamma^\mu,\gamma^\nu]$. As we can see, the effective Hamiltonian is the same as that given in Ref.~\cite{Schwinger:1951nm}; thus, ${\rm tr}\langle x|U(s)|x\rangle$,
$\langle x(s)|x'(0)\rangle$ and $\langle x(s)|\Pi|x'(0)\rangle$ can also be evaluated in the same way. Finally, the quark propagator with flavor ${\rm f}=u,d$ can be given explicitly as:
\begin{widetext}
\begin{eqnarray}
{\cal S}_{\rm f}(x,x')&\equiv&e^{-iq_{\rm f}\int_{x'}^xA_\mu dx^\mu}\hat{\cal S}_{\rm f}(x-x')\nonumber\\
\hat{\cal S}_{\rm f}(x-x')&=&{-i\over(4\pi)^2}\int_0^\infty {ds\over s^2}{-(q_{\rm f}s)^2I_2\over\text{Im}\cosh\big(iq_{\rm f}s(I_1+2iI_2)^{1/2}\big)}
	\big[-{1\over2}\gamma\big(q_{\rm f}F\coth(q_{\rm f}Fs)+q_{\rm f}F\big)(x-x')+m-{\rm sgn}(q_{\rm f})i\gamma^5\pi^0\big]\nonumber\\
	&&\exp\Big\{-iM^2s+{i\over4}(x-x')q_{\rm f}F\coth(q_{\rm f}Fs)(x-x')+{i\over2}q_{\rm f}\sigma Fs\Big\},
\end{eqnarray}
where all the Lorentz indices are suppressed but the Einstein summation rule should be understood, e.g. $\sigma F\equiv \sigma_{\mu\nu}F_{\mu\nu}$, the tensor $\coth(q_{\rm f}Fs)$ should be understood in the Taylor expansion series, $\hat{\cal S}_{\rm f}(x-x')$ is the effective quark propagator without Schwinger phase and $I_1=B^2-E^2$ is the first Lorentz invariant for EM field~\cite{Schwinger:1951nm,Cao:2015dya}. This formula is general for any configurations of constant EM field with $\pi^0$ condensate. For the recent study with $\bf{E\parallel B}$, we have
	\begin{eqnarray}
	\exp({i\over2}q_{\rm f}\sigma Fs)
	=\cos(q_{\rm f}Bs)\cosh(q_{\rm f}Es)+i\sin(q_{\rm f}Bs)\sinh(q_{\rm f}Es)\gamma^5
	+\sin(q_{\rm f}Bs)\cosh(q_{\rm f}Es)\gamma^1\gamma^2+i\cos(q_{\rm f}Bs)\sinh(q_{\rm f}Es)\gamma^4\gamma^3.
	\end{eqnarray}
It is usually more convenient and useful to transform the effective propagator to energy-momentum space. Then, by taking Fourier transformation and variable transformation $s
\rightarrow -is$, the effective quark propagator becomes:
	\begin{eqnarray}\label{propagator}
	\hat{\cal S}_{\rm f}({p})&=&\int_0^\infty {ds}\exp\big\{-M^2s-{\tan(q_{\rm f}Es)\over q_{\rm f}E}({p}_4^2+p_3^2)-{\tanh(q_{\rm f}Bs)\over q_{\rm f}B}(p_1^2+p_2^2)\Big\}\big[m-{\rm sgn}(q_{\rm f})i\gamma^5\pi^0\nonumber\\
	&&-\gamma^4(p_4-{\tan(q_{\rm f}Es)}p_3)-\gamma^3(p_3+{\tan(q_{\rm f}Es)}p_4)-\gamma^2(p_2-i~{\tanh(q_{\rm f}Bs)}p_1)-\gamma^1(p_1+i~{\tanh(q_{\rm f}Bs)}p_2)\big]\nonumber\\
	&&\Big[1-{i~\tanh(q_{\rm f}Bs)\tan(q_{\rm f}Es)\gamma^5}
	-i~\tanh(q_{\rm f}Bs){\gamma^1\gamma^2}+\tan(q_{\rm f}Es){\gamma^4\gamma^3}\Big].
	\end{eqnarray}
	
Thus, the explicit forms of the gap equations follow straightforwardly by substituting the quark propagators into Eq.(\ref{gap1}) and Eq.(\ref{gap2}):
\begin{eqnarray}
{m-m_0\over 2G}&=&m~F(M)+m\sum_{f=u,d}{N_c\over4\pi^2}\int_0^\infty {ds\over s^2}e^{-M^2s}\Big[{q_{\rm f}Es
	\over\tan(q_{\rm f}Es)}{q_{\rm f}Bs
	\over\tanh(q_{\rm f}Bs)}-1\Big]-{N_c\over4\pi^{2}}{\pi^0\over{M^2}}
(q_{\rm u}^2-q_{\rm d}^2)EB,\label{mgap}\\
{\pi^0\over 2G}&=&\pi^0F(M)+\pi^0\sum_{f=u,d}{N_c\over4\pi^2}\int_0^\infty {ds\over s^2}e^{-M^2s}\Big[{q_{\rm f}Es
	\over\tan(q_{\rm f}Es)}{q_{\rm f}Bs
	\over\tanh(q_{\rm f}Bs)}-1\Big]+{N_c\over4\pi^{2}}{m\over{M^2}}
(q_{\rm u}^2-q_{\rm d}^2)EB,\label{pi0gap}\\
F(M)&=&N_c{M\over\pi^2}\Big[\Lambda\Big({1+{\Lambda^2\over M^2}}\Big)^{1/2}-M\ln\Big({\Lambda\over M}
+\Big({1+{\Lambda^2\over M^2}}\Big)^{1/2}\Big)\Big],
\end{eqnarray}
where we have used the "vacuum regularization" scheme as in Ref.~\cite{Cao:2014uva,Cao:2015dya}. Finally, the thermodynamic potential can be derived consistently by combining the integration over $m$ of Eq.(\ref{mgap}) and  the integration over $\pi^0$ of Eq.(\ref{pi0gap}):
\begin{eqnarray}
\Omega&=&{(m-m_0)^2+(\pi^0)^2\over 4G}-{N_c M^3\over4\pi^2}\left[\Lambda\Big(1+{2\Lambda^2\over M^2}\Big)\sqrt{1+{\Lambda^2\over M^2}}-M\ln\left({\Lambda\over M}
+\sqrt{1+{\Lambda^2\over M^2}}\right)\right]+N_c\sum_{f=u,d}{1\over8\pi^2}\int_0^\infty {ds\over s^3}e^{-M^2s}\nonumber\\
&&\Big[{q_{\rm f}Es
	\over\tan(q_{\rm f}Es)}{q_{\rm f}Bs
	\over\tanh(q_{\rm f}Bs)}-1\Big]{-{N_c\over4\pi^{2}}\tan^{-1}\Big({\pi^0\over m}\Big)
	(q_{\rm u}^2-q_{\rm d}^2)EB.} 
\end{eqnarray}

%%%%%%%%%%%%%%%%%%%%%%%%%%%%%%%%%%%%%%%%%%%%%%%%%%%%%%%%%%%%%%%%%%%%%%%%%%%%%%%%%%%%%%%%%%%%%%%%%%%%%%%%%%%%%%%%%%%%%%%%%%%%%%%%%%%%%%%%%%%%%%%%%%%%%%%%%%
\subsection{Proper thermodynamic potential at finite temperature and chemical potential}\label{Tmu}
%%%%%%%%%%%%%%%%%%%%%%%%%%%%%%%%%%%%%%%%%%%%%%%%%%%%%%%%%%%%%%%%%%%%%%%%%%%%%%%%%%%%%%%%%%%%%%%%%%%%%%%%%%%%%%%%%%%%%%%%%%%%%%%%%%%%%%%%%%%%%%%%%%%%%%%%%%

By following Ref.~\cite{Cao:2015dya}, the effects of finite temperature and chemical potential can be introduced by redefining $p_4\rightarrow \omega_n+i\mu$ ($\omega_n=(2n+1)\pi T(n\in\mathbf{Z})$ is the fermion Matsubara frequency) in the propagator Eq.(\ref{propagator}) in energy-momentum space. Correspondingly, the energy integration in the $\rm Tr$ should be substituted by Matsubara frequency summation, that is, $\int dp_4\rightarrow 2\pi T\sum_{n=-\infty}^\infty$, which then alters the explicit forms of the gap equations Eq.(\ref{gap1}) and Eq.(\ref{gap2}) to
\begin{eqnarray}
{m-m_0\over 2G}&=&4N_c\sum_{f=u,d}{-i\over(4\pi)^2}\int_0^\infty {ds\over s^2}e^{-iM^2s}\Big[{m
	\over\tan(q_{\rm f}Bs)\tanh(q_{\rm f}Es)}+{\rm sgn}(q_{\rm f}){\pi^0}\Big]\vartheta_3\left({\pi\over 2}+i{\mu\over 2T},e^{-\Big|i{q_{\rm f}E\over 4\tanh(q_{\rm f}Es)T^2}\Big|}\right)(q_{\rm f}Es)(q_{\rm f}Bs),\label{mgapT}\nonumber\\
\\
{\pi^0\over 2G}&=&4N_c\sum_{f=u,d}{-i\over(4\pi)^2}\int_0^\infty {ds\over s^2}e^{-iM^2s}\Big[{{\pi^0}
	\over\tan(q_{\rm f}Bs)\tanh(q_{\rm f}Es)}-{\rm sgn}(q_{\rm f})m\Big]\vartheta_3\left({\pi\over 2}+i{\mu\over 2T},e^{-\Big|i{q_{\rm f}E\over 4\tanh(q_{\rm f}Es)T^2}\Big|}\right)(q_{\rm f}Es)(q_{\rm f}Bs).\label{pi0gapT}\nonumber\\
\end{eqnarray}
When integrating Eq.(\ref{mgapT}) over $m$ and Eq.(\ref{pi0gapT}) over $\pi^0$, we find two inconsistent results for the thermodynamic potential due to the presence of Jacobi theta function $\vartheta_3$, which can also be verified by the inequality between ${\partial\over\partial \pi^0}\text{Eq.}(\ref{mgapT})$ and ${\partial\over\partial m}\text{Eq.}(\ref{pi0gapT})$. This only means the noncommutative between the Matsubara frequency summation and the derivative with respect to either $m$ or $\pi^0$, which is very common for the chiral anomaly phenomena~\cite{Fukushima:2011jc,Son:2012zy}.

To solve the problem, we define the order parameters in the polar coordinate way, that is, $m=M\cos(\theta)$ and $\pi^0=M\sin(\theta)$ where $\theta\in[-{\pi\over2},{\pi\over2}]$ without lose of generality. Then, the formal gap equations for the chiral mass $M$ and anomalous angle $\theta$ are respectively
\begin{eqnarray}\label{gapT2}
{M-m_0\cos(\theta)\over 2G}-{1\over V_{3+1}}\text {Tr}\;{\cal S}_{A}(x)\Big[\cos(\theta)+i\gamma^5\tau_3\sin(\theta)\Big]&=&0,\\
{Mm_0\sin(\theta)\over 2G}-{M\over V_{3+1}}\text {Tr}\;{\cal S}_{A}(x)\Big[-\sin(\theta)+i\gamma^5\tau_3\cos(\theta)\Big]&=&0,
\end{eqnarray}
which have the following explicit forms
\begin{eqnarray}
{M-m_0\cos(\theta)\over 2G}&=&{N_cM\over4\pi^{2}}\sum_{f=u,d}\int_0^\infty {ds\over s^2}e^{-iM^2s}{q_{\rm f}Bs
	\over\tan(q_{\rm f}Bs)}{q_{\rm f}Es
	\over i\tanh(q_{\rm f}Es)}\vartheta_3\left({\pi\over 2}+i{\mu\over 2T},e^{-\Big|i{q_{\rm f}E\over 4\tanh(q_{\rm f}Es)T^2}\Big|}\right),\label{MgapT}\\
{Mm_0\sin(\theta)\over 2G}&=&i{N_cM^2\over4\pi^{2}}\sum_{f=u,d}q_{\rm f}Bq_{\rm f}E\int_0^\infty {ds}e^{-iM^2s}\vartheta_3\left({\pi\over 2}+i{\mu\over 2T},e^{-\Big|i{q_{\rm f}E\over 4\tanh(q_{\rm f}Es)T^2}\Big|}\right)\label{ThgapT}
\end{eqnarray}
by substituting the propagator Eq.(\ref{propagator}). Eq.(\ref{MgapT}) and Eq.(\ref{ThgapT}) still do not give a consistent thermodynamic potential, which probably indicates the noncommutative between Matsubara frequency summation and the derivative with respect to chiral anomaly parameter $\theta$. However, it is easy to see that we've separated out the chiral anomaly part solely in Eq.(\ref{ThgapT}) which involves a subtlety: The right-hand side of Eq.(\ref{ThgapT}) is just the chiral anomaly term which must originate from the ultraviolet region in energy-momentum space or the infrared region in the proper-time integral. Then, after taking the limit $s\rightarrow0$ (which is the most important region for the proper-time integration) in the Jacobi theta function $\vartheta_3$, the integration is found to be $T,\mu$ and $M$ independent:
\begin{eqnarray}\label{Thgap}
{Mm_0\sin(\theta)\over 2G}&=&{N_c\over4\pi^2}(q_{\rm u}^2-q_{\rm d}^2)EB,
\end{eqnarray}
which then gives the chiral anomaly related part for the thermodynamic potential as:
\begin{eqnarray}
\Omega_\theta={-Mm_0\cos(\theta)\over 2G}-{N_c\over4\pi^2}\theta(q_{\rm u}^2-q_{\rm d}^2)EB,
\end{eqnarray}
consistent with the one from chiral perturbation theory~\cite{Cao:2015cka}. The normal gap equation Eq.(\ref{MgapT}) can be transformed by shifting $s\rightarrow-is$ and regularized as~\cite{Cao:2014uva,Cao:2015dya}:
\begin{eqnarray}
{M-m_0\cos(\theta)\over 2G}&=&{N_c M^2\over\pi^2}\left[\Lambda\sqrt{1+{\Lambda^2\over M^2}}-M\ln\left({\Lambda\over M}
+\sqrt{1+{\Lambda^2\over M^2}}\right)\right]-{N_cM\over\pi^2}\sum_{s=\pm}\int_0^\infty p^2 dp {1\over E(p)}{2\over 1+e^{(E(p)+s\mu)/T}}+\nonumber\\
&&{N_cM\over4\pi^{2}}\sum_{f=u,d}\int_0^\infty {ds\over s^2}e^{-M^2s}\left[{q_{\rm f}Bs
	\over\tanh(q_{\rm f}Bs)}{q_{\rm f}Es
	\over \tan(q_{\rm f}Es)}\vartheta_3\left({\pi\over 2}+i{\mu\over 2T},e^{-\Big|{q_{\rm f}E\over 4\tan(q_{\rm f}Es)T^2}\Big|}\right)-\vartheta_3\left({\pi\over 2}+i{\mu\over 2T},e^{-\Big|{1\over 4sT^2}\Big|}\right)\right]
\end{eqnarray}
with the dispersion $E(p)=\sqrt{p^2+M^2}$. Then, the normal part for the thermodynamic potential is
\begin{eqnarray}
\Omega_M&=&{M^2\!-\!2Mm_0\cos(\theta)\!+\!m_0^2\over 4G}\!-\!{N_c M^3\over4\pi^2}\left[\Lambda\Big(1\!+\!{2\Lambda^2\over M^2}\Big)\sqrt{1\!+\!{\Lambda^2\over M^2}}\!-\!M\ln\left({\Lambda\over M}
\!+\!\sqrt{1\!+\!{\Lambda^2\over M^2}}\right)\right]\!-\!{2N_c\over\pi^2}T\sum_{s=\pm}\int_0^\infty\!\!\!p^2 dp \ln\Big(1\!+\!e^{-(E(p)+s\mu)/T}\Big)\nonumber\\
&&+{N_c\over8\pi^{2}}\sum_{f=u,d}\int_0^\infty {ds\over s^3}e^{-M^2s}\left[{q_{\rm f}Bs
	\over\tanh(q_{\rm f}Bs)}{q_{\rm f}Es
	\over \tan(q_{\rm f}Es)}\vartheta_3\left({\pi\over 2}+i{\mu\over 2T},e^{-\Big|{q_{\rm f}E\over 4\tan(q_{\rm f}Es)T^2}\Big|}\right)-\vartheta_3\left({\pi\over 2}+i{\mu\over 2T},e^{-\Big|{1\over 4sT^2}\Big|}\right)\right].
\end{eqnarray}
Finally, the complete thermodynamic potential can be given consistently as
\begin{eqnarray}\label{omegaT}
\Omega&=&\Omega_M-{N_c\over4\pi^2}\theta(q_{\rm u}^2-q_{\rm d}^2)I_2,\ \ \ \theta\in[-{\pi\over2},{\pi\over2}].
\end{eqnarray}
\end{widetext}

%%%%%%%%%%%%%%%%%%%%%%%%%%%%%%%%%%%%%%%%%%%%%%%%%%%%%%%%%%%%%%%%%%%%%%%%%%%%%%%%%%%%%%%%%%%%%%%%%%%%%%%%%%%%%%%%%%%%%%%%%%%%%%%%%%%%%%%%%%%%%%%%%%%%%%%%%
\section{Numerical calculations and discussions}\label{calculations}
\subsection{Zero temperature}
In order to perform numerical calculations, the three parameters of the NJL model were fixed to $G=4.93~{\rm GeV}^{-2}$, $\Lambda=0.653~{\rm GeV}$ and $m_0=5~{\rm MeV}$ by fitting the pion mass $m_\pi=134~{\rm MeV}$, pion decay constant $f_\pi=93~{\rm MeV}$ and quark condensate $\langle\bar\psi\psi\rangle=-2\times (0.25~{\rm GeV})^3$~\cite{Zhuang:1994dw}.

\begin{figure}[!htb]
	\begin{center}
		\includegraphics[width=8cm]{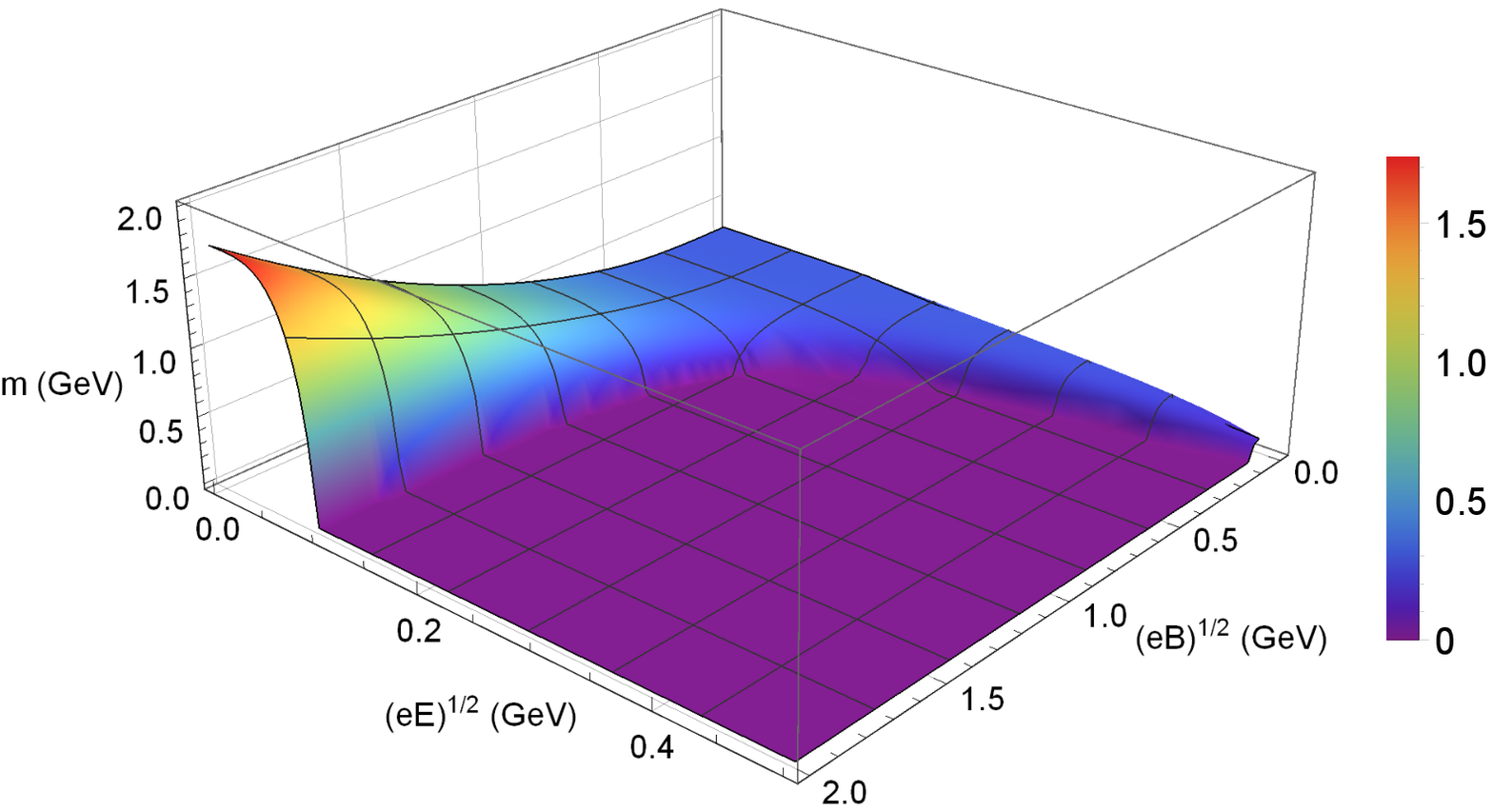}
		\includegraphics[width=8cm]{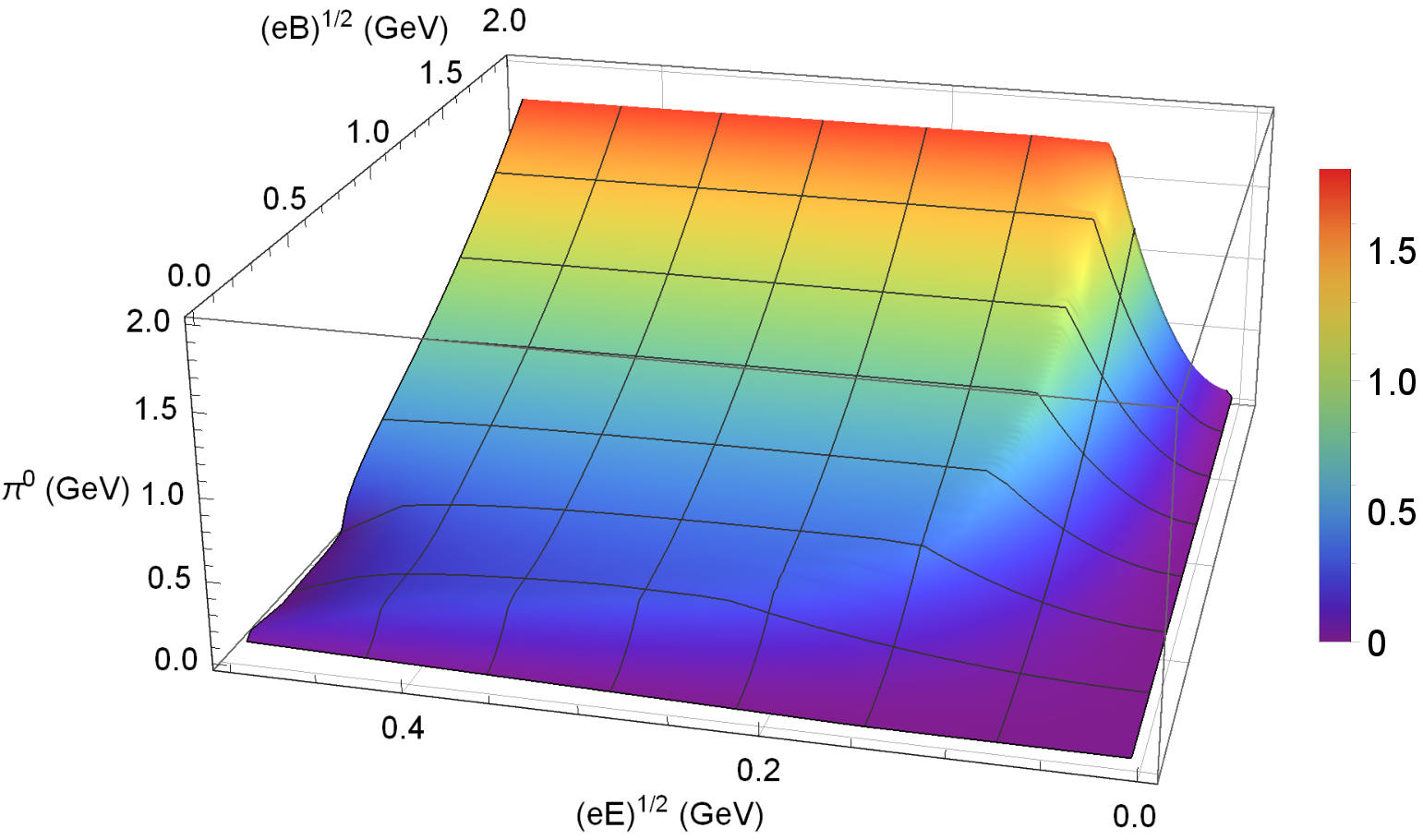}
		\caption{The mass gap $m$ and pion condensate $\pi^0$ as functions of the parallel electric field $E$ and magnetic field $B$.}\label{mpi}
	\end{center}
\end{figure}

The 3D plots of the mass gap and pion condensate with respect to the electric field $E$ and magnetic field $B$ are given in Fig.\ref{mpi}. As can be seen, in the pure electric field or magnetic field limit, the pion condensate $\pi^0$ vanishes because $I_2=0$ for these cases and the features of the mass gap $m$ are similar to those found previously~\cite{Cao:2014uva,Cao:2015xja,Cao:2015dya}. For the case with both nonzero $E$ and $B$, some interesting features are found: the mass gap $m$ keeps increasing with $B$ at small $E$, first increases and then decreases with $B$ at medium $E$ and decreases with $B$ at large $E$; the pion condensate $\pi^0$ keeps increasing with $B$ for a given $E$ but will eventually decrease with $E$ for a given $B$. It is proper to mention here that the phase transition around $m=0$ which is just the end of chiral rotation is actually of weak first-order, as will be demonstrated more explicitly in next section. Most significantly, the nontrivial behaviors of $m$ just show the competition between magnetic catalysis effect and chiral rotation effect, both induced by $B$ in the presence of parallel $E$. 

In the following, we take small electric field but large magnetic field limit to show why $m$ should vary with EM field in such a way. First, we already know a general expression for pion condensate in the parallel EM field as can be obtained from Eq.(\ref{mgap}) and Eq.(\ref{pi0gap})~\cite{Cao:2015cka}:
\begin{eqnarray}\label{pi0}
\pi^0={N_cG\over2m_0\pi^{2}}(q_{\rm u}^2-q_{\rm d}^2)EB.
\end{eqnarray}
Substituting this back into Eq.(\ref{mgap}), the gap equation for $m$ becomes 
\begin{eqnarray}\label{mgapr}
{m-{m^2\over M^2}m_0\over 2G}&=&mF(M)+mN_c\sum_{f=u,d}{1\over4\pi^2}\int_0^\infty {ds\over s^2}e^{-M^2s}\nonumber\\
&&\ \ \ \ \ \ \ \Big[{q_{\rm f}Es
	\over\tan(q_{\rm f}Es)}{q_{\rm f}Bs
	\over\tanh(q_{\rm f}Bs)}-1\Big].
\end{eqnarray}
If $m$ increases with $B$ at large magnetic field, then $m\gg m_0$. And as the contribution $F(M)$ decreases with $M$, both the $m_0$ and $F(M)$ terms can be neglected here, then the gap equation is simplified to
\begin{eqnarray}\label{gaps}
{1\over 2G}=\sum_{f=u,d}{N_c\over4\pi^2}\int_0^\infty {ds\over s^2}e^{-M^2s}\Big[{q_{\rm f}Es
	\over\tan(q_{\rm f}Es)}{q_{\rm f}Bs
	\over\tanh(q_{\rm f}Bs)}-1\Big].
\end{eqnarray}
A comment here: Usually for $M\sim \Lambda$, artifacts would be expected due to the regularization, see Ref.~\cite{He:2005nk,Cao:2015xja}; but it is not the case for large magnetic field. With increasing $B$, the second term on the right-hand side of the gap equation Eq.(\ref{mgapr}) dominates over the first one and is renormalizable even in the effective model. Thus, the qualitative results of NJL model are still credible for large magnetic field as quantum electrodynamics (QED) dominates over the four fermions interactions now. Of course, for a real QCD system, the feedback of external EM field to the effective coupling constant $G$ should also be taken into account. 

For later application, we work out the proper time integration involved in the gap equation Eq.(\ref{gaps}) in the pure magnetic field limit ($E\rightarrow 0$):
\begin{eqnarray}\label{integral}
\!\!\!&&\int_0^\infty {ds\over s^2}e^{-M^2s}\Big[{q_{\rm f}Bs
	\over\tanh(q_{\rm f}Bs)}-1\Big]\nonumber\\
\!\!\!&=&M^2\Big[1+\ln\Big({2|q_{\rm f}B|\over M^2}\Big)\Big]+|q_{\rm f}B|\ln\Big[{M^2\over4\pi|q_{\rm f}B|}\Gamma^2\Big({M^2\over2|q_{\rm f}B|}\Big)\Big].
\end{eqnarray}
The result for the pure electric field limit ($B\rightarrow0$) can just be obtained by taking the analytic continuation $|q_{\rm f}B|\rightarrow i|q_{\rm f}E|$ when $E$ is not too small. Then, by suppressing the electric field in Eq.(\ref{gaps}) and using the integral result Eq.(\ref{integral}), the gap equation can be solved to give the asymptotic form of $M$ for $B\rightarrow\infty$ (and of course $M\rightarrow\infty$):
\begin{eqnarray}\label{M}
M^2={N_cG\over6\pi^2}(q_{\rm u}^2+q_{\rm d}^2)B^2,
\end{eqnarray}
which is qualitatively consistent with that found in the LQCD calculations~\cite{Bali:2011qj,Bali:2012zg} if we remember $M\propto\langle\bar{\psi}\psi\rangle$ in the pure magnetic field case within NJL model. This also suggests that magnetic catalysis remains for chiral mass $M$. Finally, we find a critical electric field $E_c$ below which the mass $m$ keeps increasing with the magnetic field $B$ and above which $m$ eventually decreases with $B$:
\begin{eqnarray}
&&{N_cG\over2m_0\pi^2}(q_{\rm u}^2-q_{\rm d}^2)E_cB=\Big({N_cG\over6\pi^2}\Big)^{1/2}(q_{\rm u}^2+q_{\rm d}^2)^{1/2}B,\nonumber\\ &&eE_c={m_0\over3}\Big({N_cG\over6\pi^2}\Big)^{-1/2}{e(q_{\rm u}^2+q_{\rm d}^2)^{1/2}\over q_{\rm u}^2-q_{\rm d}^2}=(86.4{\rm MeV})^2.
\end{eqnarray}
The result is in very good agreement with the numerical results shown in the upper panel of Fig.\ref{mpi}.

\begin{figure}[!htb]
	\begin{center}
		\includegraphics[width=8cm]{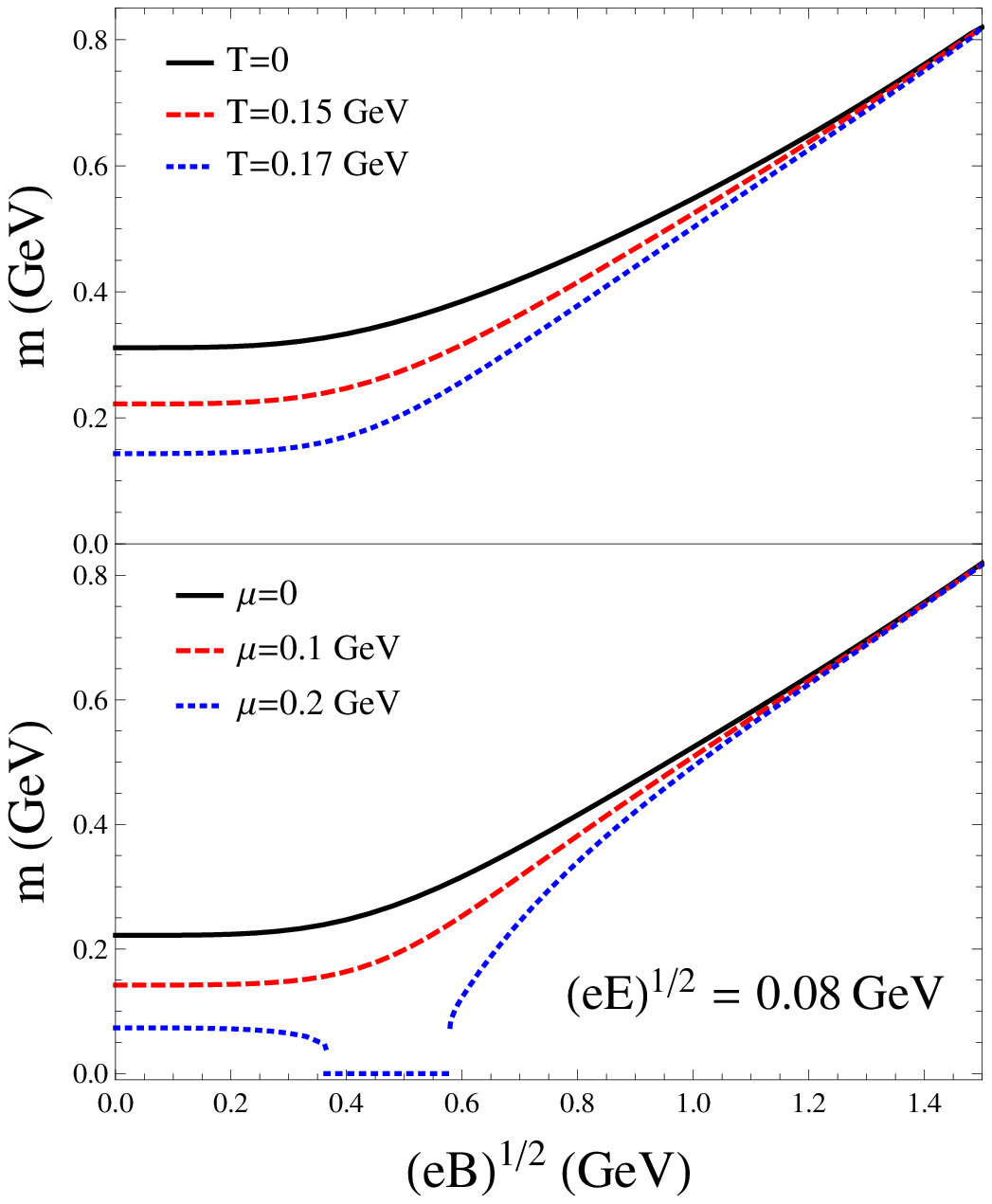}
		\caption{The mass gap $m$ as a function of magnetic field $B$ for different subcritical temperatures at vanishing chemical potential (upper panel) and for different chemical potentials at temperature $T=0.15~{\rm GeV}$ (lower panel) for the subcritical electric field case.}\label{subE}
	\end{center}
\end{figure}

\begin{figure}[!htb]
	\begin{center}
		\includegraphics[width=8cm]{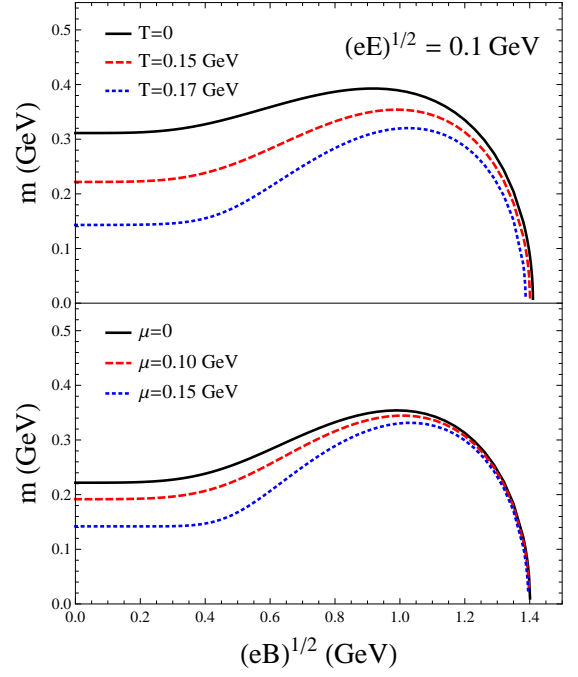}
		\caption{The mass gap $m$ as a function of magnetic field $B$ for different subcritical temperatures at vanishing chemical potential (upper panel) and for different chemical potentials at temperature $T=0.15~{\rm GeV}$ (lower panel) for the supercritical electric field case.}\label{superE}
	\end{center}
\end{figure}
%%%%%%%%%%%%%%%%%%%%%%%%%%%%%%%%%%%%%%%%%%%%%%%%%%%%%%%%%%%%%%%%%%%%%%%%%%%%%%%%%%%%%%%%%%%%%%%%%%%%%%%%%%%%%%%%%%%%%%%%%%%%%%%%%%%%%%%%%%%%%%%%%%%%%%%%%%
\subsection{Finite temperature and chemical potential}
%%%%%%%%%%%%%%%%%%%%%%%%%%%%%%%%%%%%%%%%%%%%%%%%%%%%%%%%%%%%%%%%%%%%%%%%%%%%%%%%%%%%%%%%%%%%%%%%%%%%%%%%%%%%%%%%%%%%%%%%%%%%%%%%%%%%%%%%%%%%%%%%%%%%%%%%%%

\begin{figure}[!htb]
	\begin{center}
		\includegraphics[width=8cm]{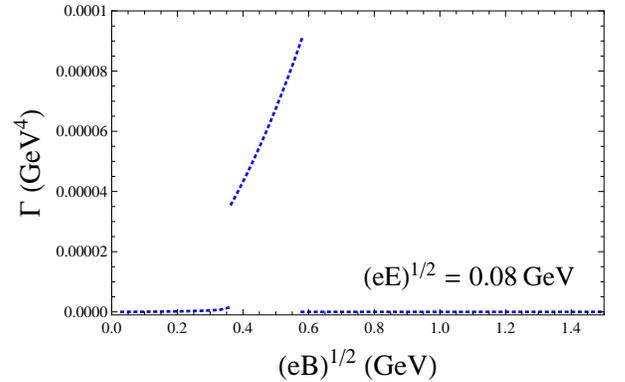}
		\caption{The pair production rate as a function of magnetic field $B$ at $T=0.15~{\rm GeV}$ and $\mu=0.2~{\rm GeV}$ for the subcritical electric field case.}\label{Gamma}
	\end{center}
\end{figure}

In order to study how the features of mass gap $m$ will be affected by finite temperature and chemical potential, we choose both a subcritical electric field $(eE)^{1/2}=0.08~{\rm GeV}$ and a supercritical electric field $(eE)^{1/2}=0.1~{\rm GeV}$ for illumination. The results are presented in Fig.\ref{subE} and Fig.\ref{superE}, respectively. There is one feature in common for both figures: the behaviors of mass gap $m$ in medium are similar to those in vacuum in the large magnetic field region, regardless of the temperature and chemical potential, specifically, they are very close to each other. This is because other parameters, such as $T$ and $\mu$, are not so important for large magnetic field, which definitely indicates the critical electric field will not change even at finite temperature and chemical potential. Besides, for proper temperature and chemical potential, such as $T=0.15~{\rm GeV}$ and $\mu=0.2~{\rm GeV}$ for the subcritical electric field case, the de Haas-van Alphen (dHvA) oscillation shows up with increasing magnetic field, similar to that found in Ref.~\cite{Cao:2015xja}. There is a region, that is $(eB)^{1/2}\in(0.364,0.579)~{\rm GeV}$, where the mass gap vanishes, which actually corresponds to the boundary minimum of the thermodynamic potential Eq.(\ref{omegaT}) with $M=0$ and $\theta={\pi\over2}$. This sudden vanishing of $M$ is very important for Schwinger pair production even at subcritical electric field. For the parallel EM field, the pair production rate will be simply modified by magnetic field as~\cite{Cohen:2008wz}:
\begin{eqnarray}
\Gamma=\left(\sum_{\rm f=u,d}{N_c(q_{\rm f}E)^2\over4\pi^3}e^{-\pi M^2/|q_{\rm f}E|}\right){\pi B/E\over\tanh(\pi B/E)},
\end{eqnarray}
which seems to increase with magnetic field $B$. However, one should remember the magnetic catalysis effect: dynamical mass usually increases with magnetic field as a power-law $M\propto (eB)^\alpha(\alpha>0)$ (see the previous section and Ref.~\cite{Miransky:2015ava}), which just means exponential suppression of the pair production rate. Thus, in order to facilitate pair production rate with magnetic field, we need some special mechanism, such as dHvA oscillation, to suppress magnetic catalysis effect. As shown in Fig.\ref{Gamma}, in the dHvA suppression region, the pair production rate is greatly enhanced and almost linearly increases with magnetic field $B$ as $\tanh(x)\approx1$ for not too small $x$.

\begin{figure}[!htb]
	\begin{center}
		\includegraphics[width=8cm]{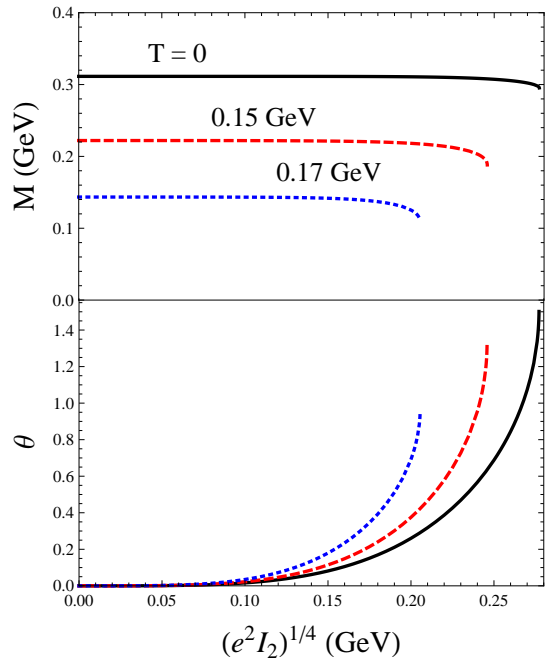}
		\caption{The chiral mass $M$ and anomalous angle $\theta$ as functions of $I_2^{1/2}=E=B$ in the chiral rotation region for different subcritical temperatures. The upper limit of $\theta$ is ${\pi\over2}$ in the lower panel.}\label{CR}
	\end{center}
\end{figure}

Finally, it is illuminative to demonstrate the chiral rotation with respect to $\sqrt{I_2}=E=B$ at finite temperature, see Fig.\ref{CR}. For brevity, we will not explore the region beyond the end of chiral rotation where $\theta={\pi\over2}$ is the boundary minimum of thermodynamic potential. At zero temperature, it can be seen explicitly in the lower panel of Fig.\ref{CR} that the transition is of weak first-order around the end of chiral rotation where $\theta$ jumps from $\lesssim{\pi\over2}$ to ${\pi\over2}$. For a larger temperature, the first-order transition is even more obvious, where the chiral rotation ends at a smaller anomalous angle than ${\pi\over2}$. This feature suggests that chiral symmetry restoration (such as that induced by temperature here) facilitates chiral rotation, which is consistent with the fact that $\theta$ is a monotonic decreasing function of $M$ from Eq.(\ref{Thgap}).

%%%%%%%%%%%%%%%%%%%%%%%%%%%%%%%%%%%%%%%%%%%%%%%%%%%%%%%%%%%%%%%%%%%%%%%%%%%%%%%%%%%%%%%%%%%%%%%%%%%%%%%%%%%%%%%%%%%%%%%%%%%%%%%%%%%%%%%%%%%%%%%%%%%%%%%%%
\section{Conclusions and prospectives}\label{conclusions}
In this paper, the competition between magnetic catalysis effect and chiral rotation effect is studied in the presence of a general parallel EM field. At zero temperature, three different features are found for the mass gap at different fixed electric fields: $m$ increases with $B$ for small $E$, increases and then decreases for medium $E$ and decreases for large $E$. This is analytically shown to be a result of the competition between MCE and CRE in the large magnetic field limit and a critical electric field $E_c$ is found which plays the role of the boundary between the small and medium electric field regions. However, the decreasing of $m$ doesn't necessarily mean inverse magnetic catalysis effect to chiral symmetry breaking: As has been stated in Ref.~\cite{Cao:2015cka}, the chiral mass $M$ is now the actual order parameter of chiral symmetry in parallel EM field, and the magnetic catalysis effect can be discovered to continue from either the analytic result Eq.~(\ref{M}) or the numerical results shown in Fig.\ref{mpi}. The fate of pion condensate after chiral rotation is also analyzed in more detail with respect to the electric field and magnetic field: While $\pi^0$ increases with $B$ for a fixed $E$, it decreases with $E$ for a fixed $B$ thanks to the Landau levels induced by $B$. 

At finite temperature $T$ and chemical potential $\mu$, a proper thermodynamic potential is found by taking into account the fact that chiral anomaly is only related to the ultraviolet dynamics in energy-momentum space. Then, the numerical calculations show that the critical electric field $E_c$ will not change with $T$ or $\mu$, because only large magnetic field region is important for the derivation of $E_c$. For large $T$ and $\mu$, the dHvA oscillation is found with respect to increasing $B$, which then greatly enhances the pair production rate even at subcritical electric field. This is important for searching Schwinger pair production of light quarks in peripheral HICs where the magnetic field is usually much larger than the electric field. In real QCD, the features might be rather different due to the inverse magnetic catalysis effect at finite temperature but this will further favor pair production. We leave the more realistic study to the future since the mechanism of IMCE is still not clear. Finally, chiral rotation effect is explored with respect to $\sqrt{I_2}=E=B$ at finite temperature, which indicates that chiral symmetry restoration usually facilitates chiral rotation.

The work can be extended by exploring the properties of collective modes for different $E$ and $B$ -- they are expected to distinguish the subcritical region from the supercritical region. Previously, the domain wall of $\pi^0$ was found to exist in the presence of both $\mu$ and $B$~\cite{Son:2007ny}, thus it is important to check if the homogeneous $\pi^0$ condensation is stable or not under the competition between $E$ and $\mu$. 

\emph{Acknowledgments}---
We thank Pengfei Zhuang from Tsinghua University and Xu-guang Huang from Fudan University for helpful discussions. LW is  is supported by the NSFC and MOST grant Nos. 11335005, 11575093, 2013CB922000 and 2014CB845400. GC is supported by the Thousand Young Talents Program of China, Shanghai Natural Science Foundation with Grant No. 14ZR1403000, NSFC with Grant No. 11535012 and No. 11675041, and China Postdoctoral Science Foundation with Grant No. KLH1512072.

\end{document}